\documentclass[a4paper,nofootinbib,twocolumn,prl,floatfix,showpacs,showkeys]{revtex4-1}

\usepackage{epsfig}
\usepackage{amsmath,amsfonts,amssymb}
\usepackage{t1enc}

\newcommand{\afb}{A_\text{FB}}
\newcommand{\ac}{A_\text{C}}

\newcommand{\bmu}{\mathcal{B}_\mu}

\newcommand{\gmu}{\mathcal{G}_\mu}

\newcommand{\omf}{\omega^4}
\newcommand{\OMf}{\Omega^4}

\newcommand{\oh}{\textstyle \frac{1}{2}}
\newcommand{\gM}{\gamma^\mu}
\newcommand{\la}{\lambda^a}
\newcommand{\tI}{\tau^I}

\begin{document}

\title{Asymmetries in $t \bar t$ production: LHC versus Tevatron}

\author{J. A. Aguilar--Saavedra, M. P\'erez--Victoria}
\affiliation{Departamento de Física Teórica y del Cosmos and CAFPE, \\
Universidad de Granada, E-18071 Granada, Spain}

\begin{abstract}
The measurement of a charge asymmetry in $t \bar t$ production at LHC constitutes more than an independent confirmation of the forward-backward asymmetry found at Tevatron. Indeed, both measurements together can be used to identify the source of the asymmetry. This is demonstrated for the case of new $Z'$, $W'$ vector bosons and colour-sextet and triplet scalars, exchanged in $t$, $u$ channels respectively, and a very heavy axigluon in the $s$ channel. In particular, current LHC measurements disfavour $Z'$, $W'$ models above the $2\sigma$ level.
\end{abstract}


\maketitle

The top quark is singled out among the other quarks by its large mass and short lifetime, making the study of its production and decay properties specially clean. Furthermore, thanks to these particular features, it can be a sensitive probe of new physics beyond the Standard Model (SM). Actually, some observations at the Fermilab Tevatron might already be a hint of new physics. The CDF and D0 Collaborations have measured the values $\afb = 0.158 \pm 0.075$~\cite{Aaltonen:2011kc}, $\afb = 0.196 \pm 0.065$~\cite{Collaboration:2011rq}, respectively, for the forward-backward (FB) asymmetry in top quark pair production. Both are
above the SM predictions, e.g. $\afb^\text{SM} = 0.051-0.089$~\cite{Ferrario:2008wm,Bernreuther:2010ny,arXiv:1107.2606,arXiv:1109.6830}. The CDF Collaboration also reports a clear enhancement of the asymmetry at high $t\bar{t}$ invariant masses, $\afb =  0.475 \pm 0.114$ for $m_{t\bar t} > 450~\text{GeV}$ (more than three standard deviations above the SM prediction $\afb^\text{SM} = 0.088-0.12$), whereas D0 does not find a statistically significant mass dependence.

On the other hand, the CMS Collaboration has recently presented a measurement of the charge asymmetry in $t \bar t$ production at the CERN Large Hadron Collider (LHC), using 1.09 fb$^{-1}$ of data~\cite{CMS}. The reported value,
$\ac = -0.016 \pm 0.030\;\text{(stat)}^{+0.010}_{-0.019} \;\text{(syst)}$, is still dominated by the statistical uncertainty, and a much better precision is expected in the near future. Systematic uncertainties are also expected to improve with a better knowledge of the detector. Clearly, the measurement of the charge asymmetry at LHC provides an independent test of the excess observed at Tevatron. We recall here that the Tevatron asymmetry $\afb$ mentioned above is defined as the relative difference (normalised to the total number) between the number of events with $\cos \theta>0$ and $\cos \theta < 0$, with $\theta$ the angle between the top quark and initial proton in the centre of mass frame. At LHC, the charge asymmetry $\ac$ measured by the CMS Collaboration is the relative difference between events with $|\eta_t| > |\eta_{\bar t}|$ and $|\eta_t| < |\eta_{\bar t}|$, with $\eta_t$ ($\eta_{\bar t}$) the pseudo-rapidity of the top (anti)quark, $\eta = -\log \tan \theta/2$, in the laboratory frame. This definition takes advantage of the larger average longitudinal boost of $t$ quarks in $pp$ collisions associated to a FB asymmetry at the parton level.

Many SM extensions have been proposed to accommodate the FB asymmetry measured by the CDF Collaboration. These models introduce new particles which can be exchanged in $s$, $t$ or $u$ channels in the processes $q \bar q \to t \bar t$, with $q=u,d$.
 While the presence of narrow $s$-channel resonances in $t \bar t$ production could be eventually spotted by an examination of the $t \bar t$ invariant mass distribution with sufficient statistics (perhaps also requiring a centre of mass energy of 14 TeV), this is more difficult for new particles exchanged in $t$ or $u$ channels, as for example new $Z'$, $W'$ vector bosons ($t$ channel), or colour-sextet and triplet scalars ($u$ channel).

In this paper we show that, combining the measurements of the charge asymmetry at LHC and the FB asymmetry at Tevatron, it is possible to discriminate among the different models, already disfavouring some of them. To arrive at this conclusion, it is necessary to go beyond the usual analyses with a few selected benchmark points, and instead scan over all the allowed values of the couplings and masses. It is also crucial to impose existing constraints from experimental data, in order to bound their range of variation.

The most obvious and robust constraints on the $t \bar t$ asymmetries result from $t \bar t$ production itself. At Tevatron, the total cross section has been precisely measured, $\sigma = 7.50 \pm 0.48$ pb~\cite{CDFtt}, which limits the possible size of new physics contributions. Total cross section measurements at LHC will not be so restrictive because $t \bar t$ production is dominated by $gg$ fusion and the systematic and theoretical uncertainties leave more room for possible departures in $q \bar q \to t \bar t$. On the other hand, the $t \bar t$ cross section at high invariant masses is sensitive to new physics and sets constraints on the masses and couplings of any new particles giving rise to the $t \bar t$ asymmetry~\cite{AguilarSaavedra:2011vw}.  Of course,
there are additional restrictions on the extra particles, as for example the production of like-sign top pairs and dijets. They are not considered here because they can be evaded in specific models~\cite{Jung:2011zv,Grinstein:2011yv,Ligeti:2011vt}.
Furthermore, we do not attempt to reproduce the $t \bar t$ invariant mass distribution at Tevatron for the models considered, as this distribution is reasonably similar to the measured one~\cite{Aaltonen:2011kc} for most of the parameter space allowed by other constraints, nor we consider the $t \bar t j$ cross section at LHC, which can be restrictive in certain parameter space regions of $t$-channel models~\cite{arXiv:1107.4364}. 
The simplified analysis presented here suffices for our purpose. Taking into account {\it only} the constraints from the Tevatron cross section and the LHC tail, we find that different SM extensions give predictions for the asymmetries corresponding to different, often disjoint regions in the $(\afb,\ac)$ plane, rendering model discrimination feasible. The inclusion of additional constraints will only shrink the allowed regions and strengthen our conclusions.

We are also conservative in the interpretation of the $t \bar t$ production limits. There are some discrepancies between different state-of-the-art predictions for the SM $t \bar t$ total cross section at Tevatron, with some results quite close to the measured one, for example $\sigma = 7.46^{+0.66}_{-0.80}$ pb~\cite{Langenfeld:2009wd}, but also significantly smaller ones, $\sigma = 6.30 \pm 0.19^{+0.31}_{-0.23}$~\cite{Ahrens:2010zv}. While the former value requires small new physics contributions  or large new amplitudes $A_\text{new} \sim -2 A_\text{SM}$, the latter allows for moderate contributions to both the cross section and the asymmetry. Thus, when requiring agreement with the Tevatron $t \bar t$ cross section we let the SM contribution be anywhere between these two values, which makes our constraints much looser (and hence the allowed regions larger) than if we stick to one of either predictions. Taking into account the uncertainties in these theoretical predictions as well as in the experimental measurement, we require in our analysis that new physics contributions to $t \bar t$ production lie inside the interval $[-0.8,1.7]$ pb. For the LHC cross section at the high-mass tail, no dedicated analysis is available yet. Still, an examination of the invariant mass distributions that have been released~\cite{CMStail} shows that large excesses over the SM prediction are already excluded. Following Refs.~\cite{AguilarSaavedra:2011vw,Delaunay:2011gv} we take the cross section for $m_{t \bar t} > 1$ TeV as a constraint, requiring that its value is at most three times the SM prediction.

All possible vector bosons and scalars contributing to $q \bar q \to t \bar t$ have been classified in Ref.~\cite{AguilarSaavedra:2011vw} according to their transformation properties under the SM gauge group $\text{SU}(3)_C \times \text{SU}(2)_L \times \text{U}(1)_Y$. There are ten possible new vector bosons and eight types of scalars, but perhaps the most interesting extensions are new colour-singlet or octet vector bosons, and colour-triplet or sextet scalars. Their transformation properties and general interaction Lagrangians with the quarks are collected in Table~\ref{tab:lagr}. We use standard notation with left-handed doublets $q_{Li}$ and right-handed singlets $u_{Ri}$, $d_{Ri}$; $\tI$ are the Pauli matrices, $\la$ the Gell-Mann matrices normalised to $\text{tr}(\la \lambda^b) = 2 \delta_{ab}$ and $\psi^c = C \bar \psi^T$, with $C$ the charge conjugation matrix. The subindices $a,b,c$ denote colour, and $\varepsilon_{abc}$ is the totally antisymmetric tensor.

\begin{table}[ht]
\begin{center}
\caption{Some vector bosons and scalar representations mediating $q \bar q \to t \bar t$.\label{tab:lagr}}
\begin{tabular}{ccl}
\hline
\hline
Label & Rep. & \multicolumn{1}{c}{Interaction Lagrangian} \\
\hline
$\bmu$ & $(1,1)_0$ 
  & $-\left( g_{ij}^q \bar q_{Li} \gM q_{Lj} 
  + g_{ij}^u \bar u_{Ri} \gM u_{Rj} \right.$ \\
  & & $\left. + g_{ij}^d \bar d_{Ri} \gM d_{Rj} \right) \bmu $ \\[1mm]
$\bmu^1$ & $(1,1)_1$ 
  & $- g_{ij} \bar d_{Ri} \gM u_{Rj} \, \bmu^{1\dagger} + \text{h.c.}$ \\[1mm]
$\gmu$ & $(8,1)_0$
  & $- \left( g_{ij}^q \bar q_{Li} \gM \frac{\la}{2} q_{Lj} 
  + g_{ij}^u \bar u_{Ri} \gM \frac{\la}{2} u_{Rj} \right.$ \\
  & & $\left. + g_{ij}^d \bar d_{Ri} \gM \frac{\la}{2} d_{Rj} \right) \mathcal{G}_\mu^a$ \\[1mm]
$\omf$ & $(3,1)_{-\frac{4}{3}}$
  & $- g_{ij} \varepsilon_{abc} \bar u_{Rib} u_{Rjc}^c \, \omega^{4a\dagger} + \text{h.c.}$ \\[1mm]
$\OMf$ & $(\bar 6,1)_{-\frac{4}{3}}$
  & $-g_{ij} \oh \left[ \bar u_{Ria} u_{Rjb}^c + 
  \bar u_{Rib} u_{Rja}^c \right] \Omega^{4ab\dagger} + \text{h.c.}$ \\
\hline
\hline
\end{tabular}
\end{center}
\end{table}

In our analysis we take five illustrative examples representing a large fraction of the models proposed in the literature to explain the $t \bar t$ asymmetry, which also involve the three possibilities of new particle exchange in the $s$, $t$ or $u$ channels.

{\it Flavour-violating $Z'$ boson} \cite{Jung:2011zv,Jung:2009jz,Barger:2010mw,Cao:2011ew,
Berger:2011ua,Bhattacherjee:2011nr}: A neutral vector boson $\bmu$ exchanged in the $t$ channel in $u \bar u \to t \bar t$. We take its $Z'tu$ couplings to be right-handed, $g_{13}^u \neq 0$, as preferred by $B$ physics constraints. Our results are independent of this choice, however. For a real $Z'$ boson the contribution to the FB and charge asymmetries is strongly constrained by the absence of like-sign top pair production~\cite{AguilarSaavedra:2011zy}. However, the relation between $tt$ and $t \bar t$ production can be evaded by placing the new boson in a complex representation of a flavour group~\cite{Jung:2011zv}.

{\it $W'$ boson} \cite{Cheung:2009ch,Cao:2010zb,Shelton:2011hq}: A charged boson $\bmu^1$ with right-handed couplings $g_{13}$ exchanged in the $t$ channel in $d \bar d \to t \bar t$. Charged bosons with left-handed couplings can also appear in $\text{SU}(2)_L$ triplets but this possibility is again disfavoured by $B$ physics constraints.

{\it Axigluon} \cite{Ferrario:2008wm,Ferrario:2009bz,Frampton:2009rk,Bai:2011ed}: A colour octet vector $\gmu$ with axial couplings $g_{ii}^q = - g_{ii}^u = - g_{ii}^d$, produced in the $s$ channel, $q \bar q \to t \bar t$. We consider this new particle to be heavy enough not to be produced on shell; otherwise its presence would generally be noticed by a bump in the $t \bar t$ invariant mass distribution~\cite{Barcelo:2011fw} and the discrimination from $t$-, $u$-channel resonances would be straightforward. The exception to this rule is given by colour octets {\it below} the $t \bar t$ production threshold~\cite{AguilarSaavedra:2011ci} or very broad~\cite{AguilarSaavedra:2011ci,Barcelo:2011vk} but in those cases the predictions are very model-dependent and deserve a separate study~\cite{AguilarSaavedra:2011ci}. We assume the axigluon is only produced in $u\bar u$ and $d \bar d$ initial states, which give the largest fraction of the $t \bar t$ cross section at the Tevatron and the LHC, and neglect additional contributions, for example from $s \bar s$ annihilation. In any case, including these small contributions hardly affects our results. No assumptions are necessary about the relative size of first- and third-generation couplings, since the axigluon is taken heavy and the cross section is proportional to the product of couplings. (Other models~\cite{Djouadi:2009nb,Delaunay:2010dw,Burdman:2010gr,Alvarez:2010js} in which the couplings of the new colour octet are not purely axial give very similar results for the relation between $\ac$ and $\afb$, but the asymmetries generated are smaller relative to the increase in cross section.)

{\it Colour-triplet scalar}~\cite{Shu:2009xf,Arhrib:2009hu,Dorsner:2009mq,Ligeti:2011vt}: A colour triplet $\omf$ with flavour-violating $tu$ couplings $g_{13}$, necessarily right-handed, exchanged in the $u$ channel in $u \bar u \to t \bar t$. Notice that the antisymmetry in colour indices implies that diagonal couplings to $uu$, $tt$ identically vanish. 

{\it Colour-sextet scalar}~\cite{Shu:2009xf,Arhrib:2009hu,Dorsner:2009mq,Ligeti:2011vt,Grinstein:2011yv,Patel:2011eh}: A colour sextet $\OMf$, also with right-handed flavour-violating $tu$ couplings $g_{13}$, and exchanged in the $u$ channel. In contrast with $\omf$, for the sextet there may be diagonal $uu$, $tt$ couplings, albeit not related to the flavour-violating ones. They can potentially give rise to large (unobserved) $tt$ signals unless suppressed by some flavour symmetry~\cite{Grinstein:2011yv,Ligeti:2011vt}.

\begin{figure*}[htb]
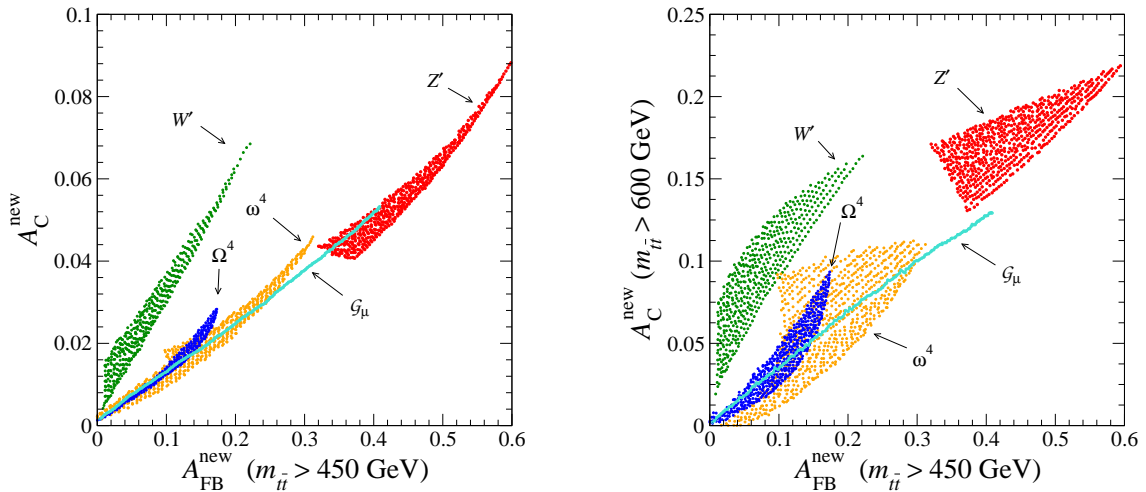

\begin{center}
\begin{tabular}{ccc}
\epsfig{file=fig1a.eps,width=6.8cm,clip=} & \quad\quad\quad\quad &
\epsfig{file=fig1b.eps,width=6.8cm,clip=}
\end{tabular}
\caption{Left: allowed regions for the new physics contributions to the FB asymmetry at Tevatron and the inclusive charge asymmetry at LHC. Right: the same, with the charge asymmetry for $m_{t \bar t} > 600$ GeV.}
\label{fig:afb}
\end{center}
\end{figure*}

The predictions of these models for the asymmetries $\afb$ and $\ac$ are found by performing a comprehensive scan over the allowed parameter space, with particle masses between 100 GeV and 10 TeV, except for the axigluon, which is assumed to be very heavy and its amplitude replaced by a four-fermion interaction~\cite{delAguila:2010mx}. The interval of the scan is adjusted as necessary to obtain a smooth variation of the predictions with the mass. The couplings are scanned uniformly in the range allowed by the Tevatron cross section limits, {\it i.e.} requiring $\Delta \sigma = [-0.8,1.7]$ pb. This constraint fixes the maximum size of the coupling for each mass considered. (The resulting allowed range for the coupling may be a single interval or the union of two, due to the competition between the interference and quadratic contributions to the cross section.) The total number of parameter space points sampled ranges between more than 2000 for the $Z'$ boson to almost 10000 for $\omf$. Our computations are performed by including the new particles and four-fermion interactions in the leading-order generator {\sc Protos}~\cite{AguilarSaavedra:2008gt}.

The new physics contributions to $\afb$ (for $m_{t \bar t} > 450$ GeV) and $\ac$ (inclusive) are presented in Fig.~\ref{fig:afb} (left), for the five models studied, taking into account the constraints on the $t \bar t$ cross section and tail mentioned above.
We only show the regions where $\afb^\text{new}$ is positive, as is the excess found by the CDF and D0 Collaborations. To a good approximation, the total asymmetries $\afb$, $\ac$ are obtained by summing the SM contributions, $\afb^\text{SM} = 0.088 \pm 0.013$~\cite{Campbell:1999ah}, $\ac^\text{SM} = 0.0130 \pm 0.0011$~\cite{Ferrario:2008wm}, to $\afb^\text{new}$ and $\ac^\text{new}$, respectively. This amounts to considering the 
dominant contributions to the total asymmetry, which are: (i) interference between new physics and tree-level SM contributions, as well as purely new physics ones; (ii) the interference between the next-to-leading order and tree-level SM. As one can observe, current LHC data already bring interesting implications for the models discussed. A salient feature of our analysis is that for the $Z'$ boson the positive asymmetries (which require a large coupling) have minimal values $\afb^\text{new} \geq 0.32$, $\ac^\text{new} \geq 0.04$ allowed by $t \bar t$ cross section constraints. Hence, the present LHC measurement of $\ac$ disfavours this model at 2.2$\sigma$ (97\% confidence level). The same measurement also disfavours the $W'$ at $2\sigma$ (95\% confidence level) if the new physics contribution to the Tevatron asymmetry is moderate, $\afb^\text{new} \geq 0.12$, as it is preferred by the CDF measurement, $\afb^\text{new} = 0.387 \pm 0.115$, and also hinted by the more recent one by the D0 Collaboration. The rest of models predict smaller asymmetries at LHC, and are less constrained by the present measurement of $\ac$. Notice that the difference between a $W'$ boson and the scalar and axigluon models stems from the different $u \bar u$ and $d \bar d$ parton densities. At Tevatron ($p \bar p$ collider) both $u,d$ from the proton and $\bar u,\bar d$ from the antiproton are valence quarks, so that $d \bar d$ is roughly $1/4$ smaller than $u \bar u$. At LHC ($pp$) both $\bar u,\bar d$ are sea quarks and $d \bar d$ is only $1/2$ smaller than $u \bar u$, resulting in a slope twice larger for the $W'$ allowed region.

Further discrimination can be achieved by the measurement of $\ac$ at high invariant masses, for example $m_{t \bar t} > 600$ GeV for which the SM cross section is only six times smaller than the total rate and statistics will be good. The result is shown in Fig.~\ref{fig:afb} (right). For a $Z'$ boson exchanged in the $t$ channel the asymmetry enhancement is much more pronounced than for the rest of models, and an unfolded measurement at high mass can definitely probe this model. (The same comment applies to $W'$ bosons.) Moreover, although apparently the scalars and the axigluon have similar predictions also for high $m_{t \bar t}$, the fact is that model parameters giving close $(\afb^\text{new},\ac^\text{new})$ points in the left-hand plot correspond to different $\ac^\text{new}$ in the right-hand one. This can be understood by recalling that for light $\omf/\OMf$ scalars  exchanged in $u$-channel the top quarks are preferrably produced in the direction of the initial {\it antiquark}. A positive $\afb$ at Tevatron can only be generated for scalar masses above few hundreds of GeV, so that the enhancement of the $u$-channel propagator in the backward direction is less pronounced. For this reason, $\ac$ at LHC is small for large $m_{t \bar t}$ except when $\omf/\OMf$ are heavy and it even decreases with $m_{t \bar t}$ for light scalars and/or high $m_{t \bar t}$.

\begin{figure}[t]
\begin{center}
\epsfig{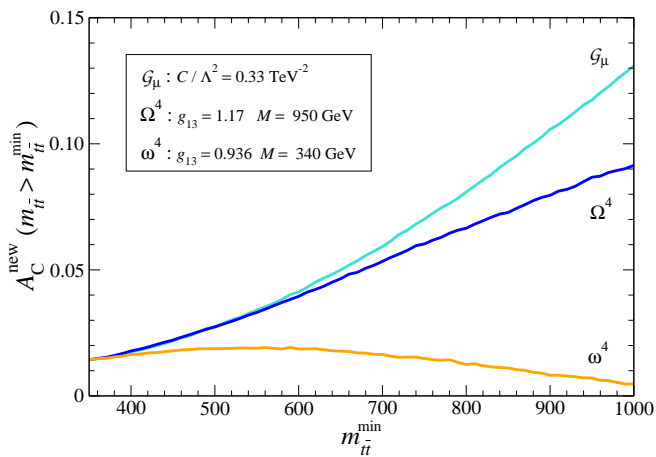} 
\caption{Dependence of the charge asymmetry on the $m_{t \bar t}$ cut, for a point with $\afb^\text{new} \simeq 0.13$, $\ac^\text{new} \simeq 0.016$ (inclusive).}
\label{fig:afb-mtt}
\end{center}
\end{figure}

These arguments provide a strong motivation for the analysis of the $m_{t \bar t}$ dependence of the charge asymmetry at LHC. To demonstrate its relevance we select one point from Fig.~\ref{fig:afb} (left), $\afb^\text{new} \simeq 0.13$, $\ac^\text{new} \simeq 0.016$ and three models yielding these values: (i) a heavy axigluon~\cite{Ferrario:2008wm,Ferrario:2009bz,Frampton:2009rk,Bai:2011ed} (see Ref.~\cite{AguilarSaavedra:2011vw} for details on the effective operators); (ii) a colour sextet~\cite{Shu:2009xf,Arhrib:2009hu,Dorsner:2009mq,Ligeti:2011vt,Grinstein:2011yv,Patel:2011eh}; (iii) a colour triplet\cite{Shu:2009xf,Arhrib:2009hu,Dorsner:2009mq,Ligeti:2011vt}. We plot in Fig.~\ref{fig:afb-mtt} the charge asymmetry as a function of the cut $m_{t \bar t}^\text{min}$ on the $t \bar t$ invariant mass.
The differences in the behaviour are striking and illustrate the general trend. In order to reproduce the same values for $\afb^\text{new}$, $\ac^\text{new}$, the colour sextet and triplet must have different mass and coupling, because the interference with the SM has opposite sign~\cite{AguilarSaavedra:2011vw,Arhrib:2009hu}. But it is the mass that mainly determines the variation of the asymmetry with $m_{t \bar t}^\text{min}$. As we have mentioned aboce, for a relatively light $u$-channel particle (for instance the $\omf$ benchmark in Fig.~\ref{fig:afb-mtt}) the asymmetry does not grow with $m_{t \bar t}$ due to the effect of the $u$-channel propagator which prefers {\it backward} top quarks, while the numerator prefers forward top quarks. For a heavier $u$-channel particle (as for example the  $\OMf$ benchmark in the same figure) the $u$-channel propagator effect is attenuated and the asymmetry reaches higher values. Thus, a more precise measurement of the inclusive charge asymmetry and an unfolded measurement at high mass will be of great help discriminating these models. Additional information can eventually be obtained from more subtle observables, such as the polarisation of the $t \bar t$ pair~\cite{Cao:2010nw}. Besides, we have also checked that for a central charge asymmetry with $|\eta_{t,\bar t}| \leq 1$~\cite{Ferrario:2008wm,Hewett:2011wz} the results are quite similar, while for a forward one some discrimination power is lost.

The allowed regions in Fig.~\ref{fig:afb} have been obtained, as explained above, by imposing a ``minimal'' set of constraints: the $t \bar t$ cross section at the Tevatron and the high invariant mass cross section at the LHC. Hence, these allowed regions contain all the possible predictions for the asymmetries in viable models.\footnote{Notice that the bulk contribution to the total cross section at Tevatron comes from the region with $m_{t \bar t} \sim 400-500$ GeV, where detection efficiency of the new physics is not very different from that of SM $t \bar t$ production. At LHC the efficiency lose at the tail for light $t$-channel particles is not very pronounced~\cite{AguilarSaavedra:2011vw}, and in any case the agreement between the SM prediction and the experimental measurement suggests much more stringent limits than the ones considered here.}
Additional constraints could be imposed, for example the $t \bar tj$ cross section at the LHC, which is important for a certain range of masses in $Z'$, $W'$ models~\cite{arXiv:1107.4364}, or the $t \bar t$ tail at the Tevatron. Doing this is not necessary in our analysis, since the regions we obtain with our minimal constraints are already disjoint, as we have shown. At any rate, we expect that the range of predictions for viable models will not be much smaller than the allowed regions shown in Fig.~\ref{fig:afb}. For example the $t \bar tj$ cross section constraint is important only for a narrow $Z'$, $W'$ mass range above the top quark mass, where on-shell associated production, {\it e.g.} $gu \to tZ' \to t \bar t u$, is large. Also, the constraints on new physics from the Tevatron tail are loosened by the smaller detection efficiency for the new contributions~\cite{Gresham:2011pa}. Though systematic scans of the parameter space for viable models have not been performed elsewhere, some sample points studied in detail~\cite{Cheung:2009ch,Jung:2009jz,Ligeti:2011vt,Grinstein:2011yv,Gresham:2011pa} suggest that most of the parameter space allowed by our constraints gives viable models.

In summary, in this paper we have investigated the relation between the $t \bar t$ asymmetries at Tevatron and LHC. If the excess found by the CDF and D0 Collaborations corresponds to new physics, the most robust probe to investigate its origin is the study of $t \bar t$ production at LHC, searching for a charge asymmetry and an enhanced $t \bar t$ tail. We have shown how the measurements of the Tevatron and LHC asymmetries can be used to identify the source of these excesses. In particular, with present data the models with $Z'$ and $W'$ bosons are already disfavoured at the 95\% confidence level. The results presented here also provide a strong motivation for the detailed study of the $m_{t \bar t}$ dependence of the charge asymmetry at LHC, which will be possible thanks to the good statistics expected at this top quark factory.

This work has been partially supported by projects FPA2010-17915 (MICINN), FQM 101 and FQM 437 (Junta de Andaluc\'{\i}a) and CERN/FP/116397/2010 (FCT).

\end{document}